\documentclass{jfm}

\usepackage{graphicx,psfrag,color}
\usepackage{amsmath}
\usepackage{amsfonts}
\usepackage{mathrsfs}
\usepackage{amssymb}
\usepackage{bm}
\usepackage{url}
\usepackage{color}
\usepackage{natbib}
\usepackage{mdwlist}

\newcommand{\bnab}{\mbox{\boldmath$\nabla$}}
\newcommand{\bnu}{\mbox{\boldmath$\nu$}}
\newcommand{\eps}{\epsilon}

\def\half {{\textstyle{1 \over 2}}}

\newcommand{\bx}{\mathbf{x}}
\newcommand{\bhx}{\mathbf{\hat{x}}}

\newcommand{\bhz}{\mathbf{\hat{z}}}
\newcommand{\bhs}{\mathbf{\hat{s}}}
\newcommand{\bhn}{\mathbf{\hat{n}}}
\newcommand{\bu}{\mathbf{u}}

\newcommand{\opL}{{\cal L}}
\newcommand{\G}{{\cal G}}
\newcommand{\Su}{{\cal S}}
\newcommand{\C}{{\cal C}}
\newcommand{\V}{{\cal V}}

\newcommand{\Vint}{\int \! \! \! \int \! \! \! \int_{\V(\Lambda)} \!\!}
\newcommand{\Vtot}{\int \! \! \! \int \! \! \! \int_{\V(1)} \!\!}

\newcommand{\Vdual}{\int \! \! \! \int \! \! \! \int_{  \overline{\V}(\Lambda)  } \!\!}

\newcommand{\diss}{\varepsilon}
\newcommand{\beq}{\begin{equation}}
\newcommand{\eeq}{\end{equation}}

\begin{document}

\noindent
\begin{center}
{\large {\bf CORRIGENDUM}}\\[10pt]

\noindent
Energy dissipation rate limits for flow through rough channels and tidal flow across topography\\[10pt]

\noindent
{\it Journal of Fluid Mechanics}, vol. 808 (2016), pp 562-575\\[10pt]

\noindent
by R. R. KERSWELL\\[15pt]
\end{center}

\section{The problem}

The upper bound derived in Kerswell (2016) is incorrect. This is because the $I_4$ and $I_5$  integrals centred at the top boundary are {\em not} exactly analogous to their counterparts at the lower boundary counter to what is  written just under  equation (2.23)  (in Kerswell 2016).  Instead, in both these cases, full volume integrals must be included due to the roughness which scale differently with $\ell$. Specifically, the estimates for the $I_4$ and $I_5$  integrals centred at the top boundary are
\begin{align} 
I_5^{top} \, & :=\, \langle \int^1_{1-\ell} \Vint |  \mathbf{u} \cdot \nabla^2 \mathbf{a} | \,\, dV\,\,d\Lambda\, \,\rangle  \nonumber \\
& \, \leq \, \ell  \,\langle  \Vtot  |\mathbf{u} \cdot \nabla^2 \mathbf{a}| \,\, dV \rangle  - \langle \,\int^1_{1-\ell}  \Vdual |\mathbf{u} \cdot  \nabla^2 \mathbf{a}| \,d\Lambda\, \,\rangle \nonumber \\
& \leq \, \qquad O(\ell \sqrt{\eps})\hspace{2cm}+ \hspace{2cm} O(\ell^{5/2} \sqrt{\eps}), \nonumber
\end{align}
where $\overline{\V}(\Lambda):=\V(1)-\V(\Lambda)$ so the last term can be estimated as in expression (2.23), and
\begin{align} 
I_4^{top} \, & :=\, \langle \int^1_{1-\ell} \Vint | \mathbf{u} \cdot (\mathbf{u} \cdot \bnab) \mathbf{a} | \,\, dV\,\,d\Lambda\, \,\rangle  \nonumber \\
& \, \leq \, \ell  \,\langle  \Vtot  |\mathbf{u} \cdot (\mathbf{u} \cdot \bnab) \mathbf{a}| \,\, dV \rangle  - \langle \,\int^1_{1-\ell}  \Vdual |\mathbf{u} \cdot (\mathbf{u} \cdot \bnab) \mathbf{a}| \,d\Lambda\, \,\rangle \nonumber \\
& \leq \, \qquad O(\ell \eps)\hspace{2cm}+ \hspace{2cm} O(\ell^3 \eps) \nonumber
\end{align}
where again the last term can be estimated as in expression (2.22). The first term of $O(\ell \eps)$ on the right hand side in this modified estimate of $I_4^{top}$ is the key new addition which breaks the bound. To see this, it is best to start with  equation (2.16) rearranged slightly as  follows
\begin{align}
(1-\ell) Gr &+\frac{1}{A} \langle
\Vtot \mathbf{u} \cdot (\mathbf{u} \cdot \bnab) \mathbf{a} + \mathbf{u} \cdot \nabla^2 \mathbf{a} \,\, dV\rangle = \nonumber \\
 &\,\frac{1}{A}  \langle \, \,\,
\frac{1}{\ell} \int^1_{1-\ell} \int_{\Su(\Lambda)} 
(\mathbf{a} \cdot \mathbf{u}) \mathbf{u} \cdot \bhn  
+ \mathbf{u} \cdot ( \bhn \cdot \bnab ) \mathbf{a} 
- \mathbf{a} \cdot ( \bhn \cdot \bnab ) \mathbf{u} \,\, dS \, d\Lambda
\nonumber\\ 
-&\frac{1}{\ell} \int^\ell_0 \int_{\Su(\Lambda)} 
(\mathbf{a} \cdot \mathbf{u}) \mathbf{u} \cdot \bhn  
+ \mathbf{u} \cdot ( \bhn \cdot \bnab ) \mathbf{a} 
- \mathbf{a} \cdot ( \bhn \cdot \bnab ) \mathbf{u} \,\, dS \, d\Lambda
\nonumber \\
& \, \hspace{1.5cm}  +\frac{1}{\ell} \int^1_{1-\ell} \Vdual \mathbf{u} \cdot (\mathbf{u} \cdot \bnab) \mathbf{a}+ \mathbf{u} \cdot \nabla^2 \mathbf{a} \,\, dV\,\,d\Lambda\,\, \nonumber \\
& \, \hspace{1.5cm}  +\frac{1}{\ell} \int^\ell_0 \Vint \mathbf{u} \cdot (\mathbf{u} \cdot \bnab) \mathbf{a}+ \mathbf{u} \cdot \nabla^2 \mathbf{a} \,\, dV\,\,d\Lambda\,\,\,
\rangle
\label{master}
\end{align}
to highlight the full volume integrals present (now on the left). Then the arguments presented in the paper are correct to reach (2.24) which now reads
\beq
(1-\ell) Gr + \frac{1}{A} \langle \Vtot \mathbf{u} \cdot (\mathbf{u} \cdot \bnab) \mathbf{a} + \mathbf{u} \cdot \nabla^2 \mathbf{a} \,\, dV \rangle
 \, \leq \,   \frac{1}{\ell} 
\biggl\{\, (B_1 \ell^2 +B_4 \ell^3) \diss+(B_2 \ell+ B_3 +B_5 \ell^2) \sqrt{\ell \diss} \,\biggr\}.
\label{master2}
\eeq
For $\ell \rightarrow 0$, this is (using the fact that $Gr=\diss/U$ )
\beq
\biggl\{ 
\frac{1}{U}+ \frac{\langle \Vtot  \mathbf{u} \cdot (\mathbf{u} \cdot \bnab) \mathbf{a}\,\, dV \rangle}{\langle \Vtot  | \bnab \mathbf{u}|^2 \,\, dV \rangle} 
\biggr\} \diss 
 \, \leq \, 
B_1 \ell \diss+ B_3 \sqrt{\diss/\ell} + \,{\rm h.o.t.} 
\eeq
since 
\beq
\langle \Vtot \mathbf{u} \cdot \nabla^2 \mathbf{a} \,\, dV\rangle \, \leq \, O(\sqrt{A \diss}) \ll B_3 \sqrt{\diss/\ell}.
\eeq
The RHS is minimised as before by $\ell =\diss^{-1/3}$ so that
\beq
\biggl\{ 
\frac{1}{U}+ \frac{\langle \Vtot  \mathbf{u} \cdot (\mathbf{u} \cdot \bnab) \mathbf{a}\,\, dV \rangle}{\langle \Vtot  | \bnab \mathbf{u}|^2 \,\, dV \rangle} 
\biggr\} \diss 
 \, \leq \, C\diss^{2/3}
\label{fact}
\eeq
where $C$ is an $O(1)$ constant or rewriting
\beq
\diss \,  \leq \, C^3 U^3
\biggl/
\biggl\{ 
1+ \frac{U\langle \Vtot  \mathbf{u} \cdot (\mathbf{u} \cdot \bnab) \mathbf{a}\,\, dV \rangle}{\langle \Vtot  | \bnab \mathbf{u}|^2 \,\, dV \rangle} 
\biggr\}^3 \biggr. .
\eeq
Since no lower bound is available on the denominator, this does not provide a bound on $\diss$. In fact, the better way to view  (\ref{fact}) is that it  presents an upper bound on the denominator
\beq
\biggl\{ 
1+ \frac{U\langle \Vtot  \mathbf{u} \cdot (\mathbf{u} \cdot \bnab) \mathbf{a}\,\, dV \rangle}{\langle \Vtot  | \bnab \mathbf{u}|^2 \,\, dV \rangle} 
\biggr\} \biggr.
 \, \leq \, O(U\diss^{-1/3})
\eeq
rather than a bound on $\diss$.

%
%
\section{ Why there is no quick fix}

It became apparent that there must be a problem with the bound in Kerswell (2016) when a connection was very recently made \citep{C17} between the `boundary layer' method of Otto \& Seis \citep{S15}  and  the Background method \citep{DC94}. It is worthwhile illustrating this connection in the simpler context of the smooth-walled channel flow problem before giving the background velocity field corresponding to the Otto-Seis `boundary layer' bounding analysis presented in Kerswell (2016).  This background field has shears throughout the interior and so cannot ever  satisfy the spectral constraint necessary to get a bound in the Background approach. This, unfortunately, makes it clear that there is no simple fix of the flawed bound in Kerswell (2016).

%
%
We adopt Seis's (2015) notation (see his \S 4) so that if
\beq
{\bf (NS)}:=\frac{\partial \bu}{\partial t}+\bu \cdot \bnab \bu+\bnab p -\nabla^2 \bu- Gr\, \bhx
\label{channel}
\eeq
then ${\bf (NS)=0}$ and $\bnab\cdot \bu=0$ with $\bu(x,y,0)=\bu(x,y,1)={\bf 0}$ define the channel flow problem.

%
%
\subsection{The Background method}

The background method is to construct the functional
\beq
\opL[\bu,\bnu]:=\langle   \int^1_0 \overline{ |\bnab \bu|^2 } \, dz\rangle - \alpha \langle  \int^1_0 \overline{ \bnu \cdot {\bf (NS)}} \, dz\rangle
\label{starting_pt}
\eeq
where $\alpha$ is a balance (scalar) parameter (usually `a' in past work), $\bnu(\bx,t)$ is a Lagrange multiplier field, and 
\beq
\overline{(\cdot)}:=\frac{1}{L_x L_y} \int^{L_x}_0 \! \! \! \int^{L_y}_0\! \! (\cdot)\, dydx
\eeq
 is an average over the rectangle $(x,y) \in [0,L_x] \times [0,L_y]$.
The key step is to restrict the difference between $\bu$ and $\bnu$ by defining a background field $\phi(z)$ such that
\beq
\bu(\bx,t)=\phi(z) \bhx+\bnu(\bx,t)
\eeq
where $\phi$ carries the mass flux of the flow but vanishes at the boundaries. In particular, if the energy dissipation rate is sought in terms of the mean flow $U$ rather than the imposed pressure gradient ($Gr$) then
\beq
U:=\int^1_0 \overline{\bu \cdot \bhx}\, dz=\int^1_0 \phi \, dz
\eeq
 (see eqn (2.29), Kerswell 2016). Rewriting $\opL$ in terms of $\bnu$ and $\phi$, and then  integrating by parts, the boundedness of the kinetic energy and  the fact that both $\phi$ and $\bnu$ vanish on $z=0$ and $1$ leads to the simplified expression
\beq
\opL[\bnu,\phi]=  \int^1_0 \phi^{'2}  \, dz - \langle \, \G(\bnu;\phi,\alpha) \, \rangle
\eeq
where 
\beq
\G:= \int^1_0  (\alpha-1)\overline{|\bnab \bnu|^2}+\overline{\alpha \nu_1 \nu_3 \phi^{'}}- (\alpha-2)\overline{ \nu_1 \phi^{''}}\, dz.
\eeq
Then the background method \citep{DC94} is the observation that 
\beq
\opL \, \leq\, \int^1_0 \phi^{'2}  \, dz - \min_{\bnu} \G(\bnu; \phi, \alpha)
\eeq
where only steady fields now need to be considered. The important point is $\min \G$ only exists for $\alpha>1$ and $\phi$ which satisfy the spectral constraint \citep{DC94}. 
The best bound is found by then minimising the whole RHS over the (convex) set of such $\phi$ and $\alpha$.

\subsection{The Otto-Seis `boundary layer' method}
%
%
The starting point for the Otto-Seis `boundary layer' approach is again (\ref{starting_pt}) and the same decomposition $\bu=\phi(z) \bhx+\bnu$ is used. The key difference now is that $\opL$ is re-expressed in terms of $\bu$ and $\phi$ rather than $\bnu$ and $\phi$. So
\begin{align}
\opL[\bu,\phi] &=\langle   \int^1_0 \overline{|\bnab \bu|^2}  \, dz\rangle - \alpha \langle  \int^1_0  \overline{(\bu-\phi(z)\bhx) \cdot {\bf (NS)}} \, dz\rangle \nonumber\\
                     &=\langle   \int^1_0  \overline{|\bnab \bu|^2}  \, dz\rangle
                     -\alpha \langle \frac{d}{dt} \int^1_0 \overline{\half \bu^2} \,dz+ \int^1_0 \overline{|\bnab \bu|^2}\, dz -Gr U \rangle \nonumber\\
                    & \hspace{4cm} -\alpha \langle \int^1_0 \phi{'}(\overline{uw}-\overline{u}_z) dz \rangle -\alpha Gr U\\
                    &=(1-\alpha)\langle   \int^1_0  \overline{|\bnab \bu|^2}  \, dz\rangle
                    -\alpha \langle  \int^1_0 \phi^{'} (\overline{uw}-\overline{u}_z) dz \rangle.
\end{align}
At this point, the Euler-Lagrange equations
\beq 
\frac{\delta \opL}{\delta \bu}={\bf 0} \quad \& \quad \frac{\delta \opL}{\delta \phi}=\frac{\delta \opL}{\delta \alpha} =0
\eeq
contain the background method  bound as {\it a} solution but there is no means to identify it as such (i.e appreciate that the associated value of $\opL$ is a bound on the dissipation rate).  Instead, the Otto-Seis approach appears to be to select a simplifying value of $\alpha=1$ so that the RHS reduces to
\beq
\diss =-\langle  \int^1_0 \phi^{'} (\overline{uw}-\overline{u}_z) dz \rangle
\label{1}
\eeq
and then to choose a simple trial function
\beq
\phi(z):= 
\frac{U}{1-\ell} \times\left\{
\begin{array}{cc}
(1-z)/\ell    & \quad    1-\ell \leq z \leq 1 \\
1               & \quad       \ell \leq z \leq 1-\ell\\
z/\ell         & \quad          0 \leq z \leq \ell\\
\end{array}
\right.
\label{phi}
\eeq
(designed so that $\int^1_0 \phi \, dz=U$ with boundary layers of size $\ell$). This converts (\ref{1}) into
\beq
(1-\ell) Gr= \frac{1}{\ell} \int^1_{1-\ell} (\overline{uw}-\overline{u}_z) \, dz
                 -\frac{1}{\ell} \int^\ell_{0} (\overline{uw}-\overline{u}_z) \, dz
\eeq
(after using $\diss=UGr$) which is eqn (4.10) in Seis (2015) and then the strategy is to bound the terms on the RHS using powers of $\diss$. The fundamental observation is that the {\em derivative}  of the background field is what appears in the Otto-Seis `boundary layer' method (Chernyshenko 2017) .\\

\subsection{Background field for the rough problem}

In the rough channel flow problem, the  background (vector) field corresponding to the `boundary layer' method as applied in Kerswell (2016)  is   $\phi ( \lambda) \mathbf{a}$ (generalised from  $\phi(z)\bhx$ in the smooth-walled problem) where
$$
\lambda:=\frac{ z-f(x,y) }{ g(x,y)-f(x,y) },  \quad \mathbf{a}:= \frac{\bhx+F_x(x,y,\lambda) \bhz}{g(x,y)-f(x,y)} \quad \& \,\,  F(x,y,\lambda):=(1-\lambda)f(x,y)+\lambda g(x,y).
$$
For example
\beq
\Vint \phi(\lambda) \mathbf{a} \cdot {\bf (NS)} \, dV = \int^1_0 \phi(\lambda) \int^{L_y}_0 \!\!\int_{\C(\lambda,y)} \bhs \cdot {\bf (NS)} \,ds \,dy\,\,\, d\lambda
\eeq
where the RHS is (2.6) of Kerswell (2016) before integrating with a general weight $\phi(\lambda)$ over $\lambda \in[0,1]$. Taking $\phi(\lambda)$ again as the piecewise-linear trial function defined in (\ref{phi}) allows the shears associated with it to be controllable. However, $\mathbf{a}$ varies spatially throughout the domain and so the shears associated with the combination do not vanish at some controlled distance from the boundary. This is what prevents the background method working and also has to break the boundary layer method. Unfortunately, this is a known limitation  of the background method with no  work-around currently on the horizon.

%


\bibliographystyle{jfm}

\end{document}